\documentclass{article}
\usepackage{subeqn}
\usepackage{graphicx}
\usepackage{natbib}
\usepackage{amssymb}
\bibpunct{(}{)}{;}{a}{}{,}
\newcommand{\beq}{\begin{equation}}
\newcommand{\eeq}{\end{equation}}
\newcommand{\beqa}{\begin{eqnarray}}
\newcommand{\eeqa}{\end{eqnarray}}
\newcommand{\beqas}{\begin{eqnarray*}}
\newcommand{\eeqas}{\end{eqnarray*}}
\newcommand{\eps}{\varepsilon}
\newcommand{\av}[1]{\left\langle#1\right\rangle}
\newcommand{\od}[1]{{\rm d}#1} 
\newcommand{\fod}[2]{\frac{{\rm d} #1}{{\rm d} #2}}
\newcommand{\pd}{\partial} 
\newcommand{\fpd}[2]{\frac{\partial #1}{\partial #2}}
\newcommand{\fpdn}[3]{\frac{\partial^{#3} #1}{\partial #2^{#3}}}
\providecommand{\operatorname}[1]{\mathop{\mathrm{#1}}\nolimits}
\newcommand{\sech}{\operatorname{sech}}
\newcommand{\half}{\mbox{$\frac{1}{2}$}}
\newcommand{\sfrac}[2]{\mbox{$\frac{#1}{#2}$}}
\newcommand{\ii}{{\rm i}} 
\renewcommand{\ll}{\!<\!\!<\!} 
\ifx\leqslant\undefined
\else

  \renewcommand{\le}{\leqslant}
  
\fi
\newcommand{\Vb}{\bar{V}}
\newcommand{\Bh}{\hat{B}}
\newcommand{\Bhy}{\hat{B}_y}

\title
{Nonlinear magnetoacoustic waves in a cold plasma}
\author{G. ROWLANDS$^1$ and  M. A. ALLEN$^2$\footnote{corresponding author}\\
\small $^1$Department of Physics, University of Warwick, Coventry, 
CV4 7AL, UK\\
\small $^2$Physics Department, Mahidol University, Rama 6 Road, 
Bangkok 10400 Thailand\\
\small (G.Rowlands@warwick.ac.uk; frmaa@mahidol.ac.th)}
\date{}
\begin{document}
\maketitle
\begin{abstract}
The equations describing planar magnetoacoustic waves of permanent
form in a cold plasma are rewritten so as to highlight the presence of
a naturally small parameter equal to the ratio of the electron and ion
masses.  If the magnetic field is not nearly perpendicular to the
direction of wave propagation, this allows us to use a multiple-scale
expansion to demonstrate the existence and nature of nonlinear wave
solutions. Such solutions are found to have a rapid oscillation of
constant amplitude superimposed on the underlying large-scale
variation. The approximate equations for the large-scale variation are
obtained by making an adiabatic approximation and in one limit, new
explicit solitary pulse solutions are found.  In the case of a
perpendicular magnetic field, conditions for the existence of solitary
pulses are derived.  Our results are consistent with earlier studies
which were restricted to waves having a velocity close to that of
long-wavelength linear magnetoacoustic waves.
\end{abstract}
\section{Introduction}
For a plasma composed of cold electrons and a single species of cold
ions, both collisions and Landau damping can be neglected with the
result that a two-fluid model provides an accurate description
\citep{KOTW68}. Such a model is governed by the continuity and
momentum equations for electrons and ions, and Maxwell's equations. In
the study of non-relativistic hydromagnetic waves with a frequency
much less than the plasma frequency, these equations may be simplified
somewhat by neglecting the displacement current and taking the number
densities of electrons and ions to be equal, except in Poisson's
equation \citep{KKT67}.  Then taking all quantities to be independent
of $y$ and $z$ one arrives at a set of equations governing planar
hydromagnetic waves.  \citet{Ili96} integrates these to obtain the
following equations for a magnetoacoustic wave of permanent form
propagating in the $x$-direction at a constant speed $V$:
\begin{subeqnarray}
\label{e:vw_x}
\fod{v}{\xi}&=&-\frac{R_i\cos\theta}{V}nw-R_iB_z,
\\
\fod{w}{\xi}&=&\frac{R_i\cos\theta}{V}nv+R_i\Bhy
+R_i\sin\theta\,(1-n), \\
\fod{\Bhy}{\xi}&=&R_enw+\frac{R_e\cos\theta}{V}nB_z, \\
\fod{B_z}{\xi}&=&-R_env-\frac{R_e\cos\theta}{V}n\Bhy,
\end{subeqnarray}
where $\xi=x-Vt$, 
\beq
\frac{1}{n}=1-\frac{1}{2V^2}\left(\Bhy^2+2\Bhy\sin\theta+B_z^2\right),
\label{e:n}
\eeq
\beq
\Bhy=B_y-\sin\theta,
\label{e:By}
\eeq
and $\theta$ is the angle between the
equilibrium magnetic field and the $x$-axis. 
In the above, $\xi$, the ion density $n$, the $y$
and $z$ components of the ion drift velocity $v$ and $w$, and the magnetic
field $(B_x, B_y, B_z)$ are normalized, respectively, by the
characteristic length $l$, the equilibrium ion density, the Alfv\'en
velocity $V_A$, and the equilibrium magnetic field strength. In such
units the speed of a linear long-wavelength magnetoacoustic wave is unity.
The
remaining parameters are defined by $R_i=\omega_{ci}l/V_A$ and 
$R_e=\omega_{ce}l/V_A$ where $\omega_{ci}$ and $\omega_{ce}$ are the
ion and electron cyclotron frequencies, respectively.
The values of the dependent variables in the absence of a wave are
zero for $v$, $w$, $\Bhy$, and $B_z$, and unity in the case of $n$.

The results we present here originate from our observation that
equations (\ref{e:vw_x}) may be rewritten so as to include a small
parameter, $\eps$. After introducing the variable $s\equiv R_e\xi/\Vb$, where
$\Vb=V\sec\theta$, the equations take the form
\begin{subeqnarray}
\label{e:vw}
\fod{v}{s}&=&-\eps\{nw+\Vb B_z\},
\label{e:v} \\
\fod{w}{s}&=&\eps\{nv+\Vb\Bhy+V\tan\theta\;(1-n)\},
\label{e:w} \\
\fod{\Bhy}{s}&=&n\{\Vb w+B_z\},
\label{e:Bhy} \\
\fod{B_z}{s}&=&-n\{\Vb v+\Bhy\},
\label{e:Bz}
\end{subeqnarray}
where $\eps\equiv R_i/R_e$ is simply the ratio of
the electron and ion masses. With $\eps\sim10^{-3}$ or smaller, any analytical
treatment should make use of the smallness of this parameter and some
form of perturbation theory is obviously called for. Furthermore, even
a purely numerical method of solution should attempt to take advantage
of the smallness of $\eps$ as otherwise it means integrating over many
small-scale variations before the underlying variation on the large scale
takes place.

From (\ref{e:vw}) it can be seen that generally 
$v$ and $w$ change on a much slower scale than $\Bhy$ and $B_z$. It
is therefore natural to make an adiabatic approximation (see
\citet{Hak-S}) which is equivalent to equating the right-hand sides of
(\ref{e:Bhy}) and (\ref{e:Bz}) to zero. This gives
\beq
B_z=-\Vb w, \qquad \Bhy=-\Vb v.
\label{e:aa}
\eeq
These relationships can be used to eliminate $\Bhy$ and $B_z$
and it is shown in Appendix~\ref{a:aa} that all the dependent variables can be
expressed in terms of a single variable which itself satisfies a Newtonian type
energy equation with a polynomial Sagdeev potential. 
This equation is then used to show the
existence of solitary pulses and nonlinear waves on the larger
scale. However, it should be stressed that the very nature of the
adiabatic approximation as used in the above is to eliminate any
variation on the smaller scale. Furthermore, the class of
solutions to (\ref{e:vw}) is restricted, in view of relationships 
(\ref{e:aa}), which implies that at some particular value of $s$ the
corresponding values of $(B_z, w)$ and $(\Bhy, v)$ are not independent.

It is the purpose of this paper to construct a perturbation expansion
based on the smallness of $\eps$ which allows one to put the adiabatic
approximation in context and to allow for rapid oscillations on the
small scale. These effects have been studied analytically in
\citet{Ili96} and later numerically by \citet{BI98} and
\citet{BIZ02}. However, those studies were only carried out for the
case $V=1+\mu/2$ where $\mu$ is small. In the present paper, the
analytic theory based on the smallness of $\eps$, a naturally small
parameter, clearly shows the origin of all the basic features of the
solutions they obtain and is not restricted to small $\mu$.

The underlying form of the governing equations suggests that a
multiple-scale perturbation expansion is appropriate and this is
carried out in Sec.~2 where explicit equations are obtained which
describe the evolution of $v$, $w$, $\Bhy$, and $B_z$ on both the
small and large scale to lowest significant order in $\eps$.  This
analysis is only valid when $\cos\theta$ is not small. For $\theta$
close to $\pm\pi/2$ singularities develop in (\ref{e:vw}) and so an
alternative set of variables and equations derived from (\ref{e:vw_x})
must be employed, as is shown in Sec.~3. The final section summarizes
our results and the further applicability of our approach is
discussed.

\section{Multiple-scale perturbation expansion}

The perturbation expansion is based on the implication from
equations (\ref{e:vw}) that two distinct
spatial scales exist -- a large one and a small one. We formally proceed by
introducing multiple scales, $s, s_1=\eps s, s_2=\eps^2 s,\ldots$ and
treating them as independent variables so that 
\[
\fod{}{s}=\fpd{}{s}+\eps\fpd{}{s_1}+\eps^2\fpdn{}{s_2}{2}+\ldots,
\]
and allowing all dependent variables $f$ to be functions of $s, s_1,
\ldots$ and expressible in the form
\[
f(s)=f_0(s,s_1,\ldots)+\eps f_1(s,s_1,\dots)+\ldots,
\]
although in the following it is not necessary to consider the scaled 
variables beyond $s$ (the small scale) and $s_1$ (the large scale). 
For a more general introduction to
this form of perturbation theory see, for example, \citet{NM-NO} and
\citet{Row-NPSE}.  

Substituting the above form for the dependent variables into
(\ref{e:v}) and (\ref{e:w}) gives to lowest order
\[
\fpd{v_0}{s}=0, \qquad \fpd{w_0}{s}=0,
\]
and so $v_0$ and $w_0$ can only be functions of $s_1$, and not $s$. At
lowest order, (\ref{e:Bhy}) and (\ref{e:Bz}) reduce to 
\beq
\label{e:DBy0}
\fpd{\Bh_{y0}}{s}=n_0\{\Vb w_0+B_{z0}\}, \qquad 
\fpd{B_{z0}}{s}=n_0\{\Vb v_0+\Bh_{y0}\}.
\label{e:DBz0}
\eeq
Since $v_0$ and $w_0$ are independent of $s$, the solution of these
equations is
\beq
\Bh_{y0}=h_y-\Vb v_0, \qquad
B_{z0}=h_z-\Vb w_0,
\label{e:B0}
\eeq
where
\[
\fpd{h_y}{s}=n_0h_z, \qquad \fpd{h_z}{s}=-n_0 h_y.
\]
These equations have the solution
\beq
h_y=h(s_1)\sin\phi, \qquad h_z=h(s_1)\cos\phi,
\eeq
where 
\beq
\fpd{\phi}{s}=n_0
\label{e:Dphi}
\eeq
and $h(s_1)$ is an as yet to be determined function of $s_1$. Substituting the
above results into (\ref{e:n}) gives, to this approximation,
\beq
\frac{1}{n_0}=A+D(s_1)\cos\phi+E(s_1)\sin\phi
\label{e:n0}
\eeq
where
\beq
A=1+\frac{2V\tan\theta\;v_0-\Vb^2(w_0^2+v_0^2)-h^2}{2V^2}
\label{e:A}
\eeq
and
\[
D=\frac{\Vb hw_0}{V^2}, \qquad E=\frac{(\Vb v_0-\sin\theta)h}{V^2}.
\]
Thus $A$, $D$, and $E$ are functions of $s_1$ only and the $s$
variation of $n_0$ is through $\phi$ only. 
Integrating the reciprocal of (\ref{e:Dphi}) after using  
(\ref{e:n0}) to express $n_0$ in terms of $\phi$ yields
\beq
A\phi+D\sin\phi-E\cos\phi=s-\tilde{s}(s_1)
\label{e:phi}
\eeq
where $\tilde{s}$ is a function of $s_1$. A solution of
(\ref{e:phi}) in the form $\phi=\phi(s,s_1)$ is obtained
in Appendix~\ref{a:phi}.

We now proceed to next order and find using (\ref{e:v}) that 
\beq
\fpd{v_1}{s}=-\fod{v_0}{s_1}-n_0w_0+\Vb^2 w_0-\Vb h\cos\phi. \\
\label{e:Dv1}
\eeq
Since the right-hand side depends on $s$ via $\phi$ only, we replace
$\pd v_1/\pd s$ by $n_0\pd v_1/\pd \phi$. Remembering that
$v_0$, $w_0$ and $h$ are independent of $\phi$, and using
(\ref{e:n0}), (\ref{e:Dv1}) is readily integrated to give
\beqa\nonumber
v_1&=&\left(\Vb^2w_0-\fod{v_0}{s_1}\right)
(A\phi+D\sin\phi-E\cos\phi)-w_0\phi \\
&&\mbox{}+\Vb h\left(\frac{E}{4}\cos2\phi
-\frac{D}{4}(2\phi+\sin2\phi)-A\sin\phi\right)+\tilde{v}(s_1)
\label{e:pv1}
\eeqa
where $\tilde{v}(s_1)$ is an undetermined function. 
To ensure that $v_1$ is a bounded function of $s$, we must remove
terms proportional to $\phi$. After replacing $D$ by its definition
this results in the consistency condition
\beq
A\fod{v_0}{s_1}=\left(A\Vb^2-1-\frac{\Vb^2h^2}{2V^2}\right)w_0.
\label{e:Dv0}
\eeq
This is the equation
for the variation of $v_0$ on the slowly varying scale, $s_1$. Using
(\ref{e:Dv0}) to simplify (\ref{e:pv1}) leaves us with
\beq
v_1=\frac{1}{A}\left(w_0+\frac{\Vb hD}{2}\right)(D\sin\phi-E\cos\phi)
+\Vb h\left(\frac{E}{4}\cos2\phi-\frac{D}{4}\sin2\phi-A\sin\phi\right)
\label{e:v1}
\eeq
in which $\tilde{v}$ has been absorbed into $v_0$.

Similarly, to first order, (\ref{e:w}) yields
\beq
\fpd{w_1}{s}=-\fod{w_0}{s_1}+n_0v_0-\Vb^2 v_0
+\Vb h\sin\phi+V\tan\theta\;(1-n_0).
\label{e:Dw1}
\eeq
After again replacing $\pd/\pd s$ by $n_0\pd/\pd \phi$
and integrating, to obtain a bounded $w_1$ we require that
\beq
A\fod{w_0}{s_1}=\left(1-A\Vb^2+\frac{\Vb^2h^2}{2V^2}\right)v_0+
V\tan\theta\left(A-1-\frac{h^2}{2V^2}\right)
\label{e:Dw0}
\eeq
with the result that 
\beqa\nonumber
w_1&=&\frac{1}{A}\left(V\tan\theta-v_0-\frac{\Vb hE}{2}\right)
(D\sin\phi-E\cos\phi)\\ && \mbox{}
-\Vb h\left(A\cos\phi+\frac{D}{4}\cos2\phi+\frac{E}{4}\sin2\phi\right).
\label{e:w1}
\eeqa

To lowest order, the adiabatic approximation (\ref{e:aa}) is
\beq
B_{z0}=-\Vb w_0, \qquad \Bh_{y0}=-\Vb v_0.
\label{e:aa0}
\eeq
Comparison with equations (\ref{e:B0}) shows that this
approximation is equivalent to setting $h=0$.
Equations (\ref{e:Dv0}) and (\ref{e:Dw0}) then form a
complete set which can be integrated. The details are given in
Appendix~\ref{a:aa}. In particular, the existence of solitary pulses
is proven.
 
If the adiabatic approximation is not made, 
it is necessary to
obtain an equation for the variation of $h$ on the $s_1$
scale. This is achieved by considering the equations for $\Bhy$ and
$B_z$ to next order in $\eps$. From (\ref{e:Bhy}) and (\ref{e:Bz}) we
may write, respectively,
\beqas
\fpd{\Bh_{y1}}{s}+\fpd{\Bh_{y0}}{s_1}&=&
n_0\{\Vb w_1+B_{z1}\}+n_1h\cos\phi, \\ 
\fpd{B_{z1}}{s}+\fpd{B_{z0}}{s_1}&=&
-n_0\{\Vb v_1+\Bh_{y1}\}-n_1h\sin\phi. 
\eeqas
Since the variation of all coefficients with $s$ is through $\phi$, we replace
$\pd/\pd s$ by $n_0\pd/\pd\phi$ and rewrite the above equations as
\begin{subeqnarray}
\fpd{\Bh_{y1}}{\phi}+\frac{1}{n_0}\fpd{\Bh_{y0}}{s_1}&=&
\Vb w_1+B_{z1}+\frac{n_1h}{n_0}\cos\phi \label{e:By1},\\ 
\fpd{B_{z1}}{\phi}+\frac{1}{n_0}\fpd{B_{z0}}{s_1}&=&
-\Vb v_1-\Bh_{y1}-\frac{n_1h}{n_0}\sin\phi.  \label{e:Bz1}
\end{subeqnarray}
We proceed by adding (\ref{e:By1}) multiplied by $\sin\phi$ to 
(\ref{e:Bz1}) multiplied by $\cos\phi$ and then integrating from 0 to
$2\pi$. Insisting that $\Bh_{y1}$ and $B_{z1}$ are periodic functions
of $\phi$ means that 
\[
\av{\fpd{\Bh_{y1}}{\phi}\sin\phi}=-\av{\Bh_{y1}\cos\phi}, \qquad
\av{\fpd{B_{z1}}{\phi}\cos\phi}=\av{B_{z1}\sin\phi},
\]
where $\av{\cdot}$ denotes the average as $\phi$ varies from 0 to $2\pi$.
The combined equations then reduce to 
the following equation for the variation of $h$:
\beq
\av{\frac{1}{n_0}}\fod{h}{s_1}=\Vb\left\{
\fod{v_0}{s_1}\av{\frac{\sin\phi}{n_0}}+\fod{w_0}{s_1}\av{\frac{\cos\phi}{n_0}}
+\av{w_1\sin\phi}-\av{v_1\cos\phi}\right\}.
\label{e:preDb}
\eeq
Using (\ref{e:n0}), (\ref{e:v1}), and (\ref{e:w1}) we have
\[
\av{\frac{1}{n_0}}=A, \qquad
\av{\frac{\sin\phi}{n_0}}=\frac{E}{2}, \qquad
\av{\frac{\cos\phi}{n_0}}=\frac{D}{2},
\]
\[
\av{w_1\sin\phi}=\left(V\tan\theta-v_0-\frac{\Vb  hE}{2}\right)\frac{D}{2A},
\qquad
\av{v_1\cos\phi}=-\left(1+\frac{\Vb^2h^2}{2V^2}\right)\frac{Ew_0}{2A}.
\]
Then inserting the above expressions and results (\ref{e:Dv0}) and
(\ref{e:Dw0}) 
into (\ref{e:preDb}) and simplifying we find that the right-hand side
of (\ref{e:preDb}) is zero and hence that $h$ is a constant. This is
in agreement with the result obtained for $V$ close to 1 in \citet{Ili96}.  

In summary, 
we have seen that to lowest order, the ion drift velocity components
$v$ and $w$ only show large-scale variation. Fast periodic variation
occurs at the next order of approximation, as given by   
(\ref{e:v1}) and (\ref{e:w1}), but given the smallness of $\eps$, these
oscillations would be barely discernible. On
the other hand, even to lowest order, the magnetic field components
show rapid oscillations on top of the large-scale variation:  
\[
\Bh_{y0}=-\Vb v_0(s_1)+h\sin\phi(s,s_1), \qquad
B_{z0}=-\Vb w_0(s_1)+h\cos\phi(s,s_1),
\]
where, as is shown in Appendix~\ref{a:phi}, $\sin\phi$ and $\cos\phi$ are
periodic functions of the variable $S$ given by (\ref{e:S}). 

\section{The small $\cos\theta$ and $\theta=\pm\pi/2$ limits}
So far we have treated $\cos\theta$ as finite but our treatment does
not allow one to pass to the case $\cos\theta=0$ since in this limit
$\Vb$ becomes infinite. To consider this
limit we write $\theta=\pm\pi/2\mp\sqrt{\eps}\psi$ so that
$\cos\theta=\sin(\sqrt\eps\psi)\simeq\sqrt\eps\psi$. (In the remainder
of this section the upper and lower signs refer to the cases where
$\theta$ is in the neighbourhood of $\pi/2$ and $-\pi/2$, respectively.)
It is now
necessary to go back to the original equations expressed in terms of
$\xi$ rather than $s$ and define $X\equiv\sqrt{R_iR_e}\xi$. Also, to
avoid singular solutions we need to use scaled versions of the ion drift
velocities, defined by $\bar{v}=v/\sqrt\eps$ and $\bar{w}=w/\sqrt\eps$.
Then dividing (\ref{e:vw_x}) by $\sqrt{R_iR_e}$ we obtain 
\begin{subeqnarray}
\label{e:psi}
\fod{\bar{v}}{X}&=&-B_z+O(\eps\psi), \\
\fod{\bar{w}}{X}&=&\Bhy\pm1-n+O(\eps\psi^2), \\
\fod{\Bhy}{X}&=&n\bar{w}+\frac{nB_z}{V}\psi+O(\eps\psi^3), \\
\fod{B_z}{X}&=&-n\bar{v}-\frac{n\Bhy}{V}\psi+O(\eps\psi^3)
\end{subeqnarray}
with 
\beq
\frac{1}{n}=1-\frac{1}{2V^2}\left(\Bhy^2\pm2\Bhy+B_z^2\right)+O(\eps\psi^2).
\label{e:npsi}
\eeq
Any solution to the above equations will
be such that $\bar{v}\sim B$ and hence the real velocity
$v=\sqrt\eps\bar{v}$ will be small compared to $B$.

For the $\theta=\pm\pi/2$ limits, equations (\ref{e:psi}) reduce to 
\begin{subeqnarray}
\label{e:psi0}
\fod{\bar{v}}{X}=-B_z, \qquad
\fod{B_z}{X}=-n\bar{v}, \label{e:vb} \\
\fod{\bar{w}}{X}=\Bhy\pm1-n, \qquad 
\fod{\Bhy}{X}=n\bar{w}. \label{e:wb}
\end{subeqnarray}
We now define an operator $L$ by
\[
Lf\equiv \fod{}{X}\left(\frac{1}{n}\fod{f}{X}\right).
\]
Then from (\ref{e:vb}) and (\ref{e:wb}) it can be seen that, respectively, 
\begin{subeqnarray}
\label{e:LB}
LB_z&=&B_z, \label{e:LBz} \\  LB_y&=&B_y-n, \label{e:LBy}
\end{subeqnarray}
in which we have re-instated $B_y$, as given in (\ref{e:By}). 
Combining equations (\ref{e:LB}) gives
\[
B_yLB_z-B_zLB_y=B_zn.
\]
Integrating this over one period (or all $X$ if boundary
conditions permit) yields
\beq
\av{B_zn}=0,
\label{e:aBzn}
\eeq
where $\av{\cdot}$ denotes the integral over $X$.
Similarly, (\ref{e:LBz}) and (\ref{e:LBy}) imply that
\beq
\av{B_z}=0, \qquad \av{B_y}=\av{n}.
\label{e:aBzy}
\eeq
Relations (\ref{e:aBzn}) and (\ref{e:aBzy}) 
are satisfied if $n$ and $B_y$ are even functions and $B_z$ is
an odd function of $X$. 

We can demonstrate the existence of 
a non-trivial solution of (\ref{e:psi0}) by taking $B_z=0$. In this case
(\ref{e:npsi}) becomes
\[
\frac{1}{n}=\alpha(1-\beta B_y^2)
\]
where
\beq
\alpha=1+\frac{1}{2V^2}, \qquad \beta=\frac{1}{1+2V^2},
\label{e:ab}
\eeq
with the result that (\ref{e:LBy}) can be re-expressed as
\[
\alpha\fod{}{X}\left((1-\beta B_y^2)\fod{B_y}{X}\right)
=B_y-\frac{1}{\alpha(1-\beta B_y^2)}.
\]
Letting $p=\od{B_y}/\od{X}$, this can be written as the following
first-order differential equation for $p^2$:
\[
\fod{p^2}{B_y}-\left(\frac{4\beta B_y}{1-\beta B_y^2}\right)p^2=
\frac{2}{\alpha}\left(
\frac{B_y}{1-\beta B_y^2}-\frac{1}{\alpha(1-\beta B_y^2)^2}\right)
\]
which has an integrating factor of $(1-\beta B_y^2)^2$. Hence the
solution is given by
\beq
p^2=\frac{2Q}{\alpha(1-\beta B_y^2)^2}
\label{e:p}
\eeq
where
\beq
Q=\frac{B_y^2}{2}-\frac{\beta B_y^4}{4}-\frac{B_y}{\alpha}+Q_0
\label{e:Q0}
\eeq
and $Q_0$ is an integration constant.
For a solitary pulse solution, the appropriate boundary conditions are
$B_y\to\pm1$ and $Q\to0$ as $|X|\to\infty$. Using these allows us to
determine $Q_0$. We then rewrite (\ref{e:Q0}) as the following expansion in
$\Bhy$:
\beq
Q=\Bhy^2\left\{\frac{1-3\beta}{2}\mp\beta\Bhy
-\frac{\beta}{4}\Bhy^2
\right\}.
\label{e:Q}
\eeq
A necessary condition for the existence of solitary pulses is therefore
that $\beta<1/3$. In addition, from (\ref{e:p}) it can be
seen that $p$ is singular at $B_y^2=1/\beta$. Hence, for a solitary
pulse to exist,  at least one of the
two non-trivial zeros of the expression for $Q$ given in (\ref{e:Q}) 
must lie within the range $-1/\sqrt\beta\mp1<\Bhy<1/\sqrt\beta\mp1$.
The zero which is larger in magnitude never satisfies this. The
remaining zero at $\pm\sqrt{2/\beta-2}\mp2$ satisfies the condition if
$\beta>1/9$. Using (\ref{e:ab}) we can now write the
sufficient condition for the existence of a solitary pulse solution as
$1<V^2<4$.

As a check on our calculation we look at the case 
when $V=1+\mu/2$ for small positive $\mu$. Then (\ref{e:p}) reduces to 
\[
\left(\fod{\Bhy}{X}\right)^2=\frac{\Bhy^2}{4}(\Bhy\pm4)(\pm\mu-\Bhy)
\]
to lowest order in $\mu$. The above equation has the solution
\[
\Bhy=\pm\mu\sech^2\half\sqrt{\mu}(X-X_0)+O(\mu^2)
\]
where $X_0$ is an arbitrary constant. This is in agreement with the
result given in \citet{BI98}.

\section{Conclusions}

We have studied a set of magnetohydrodynamic equations for 
planar magnetoacoustic waves of permanent form propagating in a 
two-component cold plasma and, by taking advantage of the smallness of
the ratio of the electron to ion masses, have obtained a reduced set of
equations which describe the large-scale variation of the
magnetoacoustic wave solution of the full equations. Superimposed on
the large-scale variation, multiple-scale perturbation analysis indicates that
there is a rapid oscillation which is of constant amplitude in the
case of the lowest-order magnetic field components. 
These results are
consistent with the study of \citet{Ili96} which was restricted to a
narrow range of velocities. In addition, the approach expounded in
this paper puts the adiabatic approximation into its true context.

In this work we have obtained various conditions for the existence of
solitary pulses. Whether these solutions correspond to phenomena that
could occur in nature depends on whether they are stable. Linear
stability analysis of the solutions shown to exist in this paper is a
challenging problem. However, the numerical solution of the full
(time-dependent) system of equations obtained by \citet{BI98}, show
that for a range of initial conditions the solution relaxes to the
type of solution shown to exist here. This suggests that our solutions
are stable, at least to perturbations applied in the direction of
propagation.

Although the equations studied here arise from a
magnetohydrodynamics problem, the method is applicable to a more
general set of nonlinear equations where two distinct scales are
a basic feature. An advantage of the present study is that the
equations obtained on the large scale can be investigated analytically
and describe real physical processes. Although we have only looked at
cold plasmas, an exactly analogous procedure can be applied to the
case of warm plasmas, at the expense of some additional algebraic
complexity. The relevant governing equations are given in \citet{BIZ02}.


\section*{\normalsize \em Acknowledgements}

G.R. thanks Mahidol University for their hospitality during his visit.

\appendix
\section{Adiabatic approximation}
\label{a:aa}

\begin{figure}
\begin{center}
\includegraphics[width=9cm]{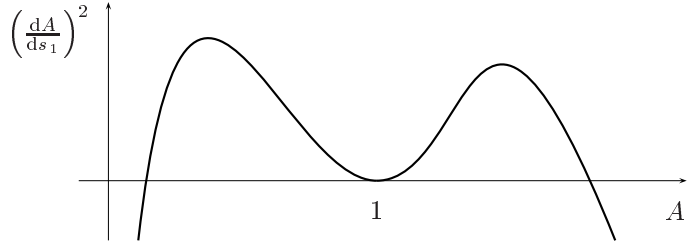}
\end{center}
\caption{Phase plane for (\ref{e:ADA2s}) when conditions are such that
  both compressive and rarefactive solitary pulses occur.}
\label{f:aapp}
\end{figure}

When $h=0$, 
the coupled equations for $v_0$ and $w_0$, namely (\ref{e:Dv0}) and
(\ref{e:Dw0}), reduce to
\beq
\fod{v_0}{s_1}=\left(\Vb^2-\frac{1}{A}\right)w_0
\label{e:aDv0}
\eeq
and 
\beq
\fod{w_0}{s_1}=\left(\frac{1}{A}-\Vb^2\right)v_0
+V\tan\theta\;\left(1-\frac{1}{A}\right).
\label{e:aDw0}
\eeq
Multiplying (\ref{e:aDv0}) and (\ref{e:aDw0}) by $2v_0$ and $2w_0$,
respectively, and adding gives
\beq
\fod{}{s_1}(v_0^2+w_0^2)=2V\tan\theta\;\left(1-\frac{1}{A}\right)w_0.
\label{e:Dv2w2}
\eeq
Differentiating (\ref{e:A}) with respect to $s_1$ 
and using the above, one obtains
\[
\fod{A}{s_1}=\frac{(\Vb^2-1)\tan\theta}{V}\frac{w_0}{A},
\]
or, provided that $\Vb^2\neq1$, 
\beq
w_0=\frac{A}{\kappa}\fod{A}{s_1},\qquad
\kappa=\frac{(\Vb^2-1)\tan\theta}{V}.
\label{e:aw0}
\eeq
Substituting (\ref{e:aw0}) into (\ref{e:Dv2w2}) and (\ref{e:aDv0}) 
and integrating yields, respectively,
\beq
v_0^2+w_0^2=C_1+V\kappa^{-1}\tan\theta(A^2-2A)
\label{e:v2w2}
\eeq
and
\beq
v_0=C_2+\kappa^{-1}(\half\Vb^2A^2-A)
\label{e:av0}
\eeq
where $C_1$ and $C_2$ are integration constants. Finally, after combining
(\ref{e:aw0}), (\ref{e:v2w2}) and (\ref{e:av0}) one obtains
\beq
\left(A\fod{A}{s_1}\right)^2=\sum_{m=0}^4\gamma_mA^m
\label{e:ADA2}
\eeq
in which 
\[
\gamma_0=(C_1-C_2^2)\kappa^2, \quad \gamma_1=2(C_2-V\tan\theta)\kappa,
\quad \gamma_2=(V\tan\theta-C_2\Vb^2)\kappa-1,
\]
\[
\gamma_3=\Vb^2, \qquad \gamma_4=-\sfrac{1}{4}\Vb^4.
\]
This is of the form of the energy equation of a particle with position
$A$ in a Sagdeev potential which is minus the right-hand side of
(\ref{e:ADA2}). In general, nonlinear waves exist and in particular
solitary pulses. The latter can occur when the boundary conditions are
such that $v\to0$, $w\to0$, and $n\to1$ as $s_1\to\pm\infty$. Using the
result that in
this limit $A\to1$, the boundary conditions allow us to use
(\ref{e:v2w2}) and (\ref{e:av0}) to determine the
integration constants in this case:
\[
C_1=V\kappa^{-1}\tan\theta, \qquad C_2=(1-\half\Vb^2)\kappa^{-1}.
\]
Then (\ref{e:ADA2}) reduces to
\beq
\left(A\fod{A}{s_1}\right)^2=\frac{(A-1)^2}{4}(a+bA+cA^2)
\label{e:ADA2s}
\eeq
where
\[
a=4(\Vb^2-1)\tan^2\theta-(\Vb^2-2)^2, \qquad
b=2\Vb^2(2-\Vb^2), \qquad c=-\Vb^4.
\]
The requirement that (\ref{e:ADA2s}) gives rise to a solitary pulse is
that $a+b+c>0$. Using the above renders this condition as
$(\Vb^2-1)(1-V^2)>0$
which on rearranging yields
\[
\cos^2\theta<V^2<1.
\]
As illustrated in Fig.~\ref{f:aapp}, since $c<0$, if the above
condition is satisfied, compressive solitary pulses will always
occur. However, rarefactive pulses are only possible if the smaller root of
$a+bA+cA^2=0$ is above zero. This will occur if $a<0$ and
$\Vb^2<2$. These requirements are equivalent to the condition
\[
V^2<2(1-|\sin\theta\,|).
\]
This implies that if the values of $V$ and $\theta$ are 
such that compressive pulses exist, then rarefactive pulses will also
occur if $|\theta|\le\pi/6$. 

It is possible to integrate (\ref{e:ADA2s}) to obtain the spatial
variation of the solitary pulses implicitly. An approximate explicit
solution can be obtained when $\Vb^2$ is just above 1. Introducing
$U\equiv A-1$, (\ref{e:ADA2s}) becomes
\beq
\left(\fod{U}{s_1}\right)^2=\frac{\Vb^4U^2(U_+-U)(U-U_-)}{(1+U)^2}
\label{e:DU}
\eeq
where
\[
U_\pm=\frac{2\sqrt{\Vb^2-1}}{\Vb^2}\left(\pm\tan\theta-\sqrt{\Vb^2-1}\right).
\]
If $\sqrt{\Vb^2-1}\ll|\tan\theta\,|$, then $U_\pm\simeq\pm\nu$ where
\[
\nu=\frac{2\sqrt{\Vb^2-1}\tan\theta}{\Vb^2}.
\]
For solitary pulse solutions, $|U|<|U_\pm|$, and so if $\tan\theta$
is of order unity, $U\ll1$. Hence (\ref{e:DU}) reduces to 
\[
\fod{U}{s_1}=U\sqrt{\nu^2-U^2}
\]
at lowest order and one obtains
\beq
A\simeq1\pm\nu\sech \nu s_1.
\label{e:Anu}
\eeq
Using (\ref{e:aw0}) and (\ref{e:av0}) we can then obtain the
corresponding expressions for $w_0$ and $v_0$:
\beq
w_0\simeq\mp4V\tan\theta\sech\nu s_1\tanh\nu s_1, \qquad
v_0\simeq2V\tan\theta\sech^2\nu s_1.
\label{e:wvnu}
\eeq
In this adiabatic approximation, the lowest order components of the
magnetic field are just multiples of these quantities, as given by
(\ref{e:aa0}). 

We can now also use (\ref{e:Anu}) and (\ref{e:wvnu}) to obtain the
solution when $\Vb^2=1$. This corresponds to the limit $\nu\to0$ in
which case $A\to1$, $w_0\to0$, and $v_0\to2V\tan\theta$. As a check on
our calculation, we note that these results are consistent with the
definition of $A$ as given by (\ref{e:A}) when $h=0$.

\section{Explicit expression for $\phi(s,s_1)$}
\label{a:phi}
The variation of $\phi$ with $s$ is given implicitly by
(\ref{e:phi}). An explicit expression can be obtained by writing
the equation in the form
\beq
S+\psi=\phi+\psi-\sigma\sin(\phi+\psi)
\label{e:impphi}
\eeq
where $S=(s-\tilde{s})/A$, 
$\psi=\arg(-D+\ii E)$ and 
$\sigma=\sqrt{D^2+E^2}/A$.
The following 
explicit solution to (\ref{e:impphi}) was first obtained by \citet{Jac60}
(although for a more transparent exposition see
p.154 of \citet{IR-NWSC}):
\beq
\phi=S+2\sum_{m=1}^\infty\frac{J_m(m\sigma)}{m}\sin m(S+\psi).
\label{e:phisum}
\eeq
It can be seen that $\phi$ has a directed component, $S$, 
on which a periodic variation is superimposed.
Since $A$ and hence the period vary on the $s_1$ timescale, 
to the order in $\eps$ to which (\ref{e:phisum}) applies, it is 
more appropriate to re-define $S$ by 
\beq
S=\int_{\tilde{s}}^s\frac{\od{s'}}{A(\eps s')}.
\label{e:S}
\eeq
Such a definition avoids secular terms at higher order in the $\eps$ expansion.

\end{document}